\documentstyle[prb,aps,preprint]{revtex}
\begin{document}
\draft
\title{Phonons and specific heat of linear dense phases of atoms
physisorbed in the grooves of carbon nanotube bundles}

\author{Antonio \v{S}iber\thanks{E-mail: asiber@ifs.hr}}
\address{Institute of Physics, Bijeni\v{c}ka c. 46, P.O. Box 304, 10001
Zagreb, Croatia}
\maketitle

\begin{center}
{\bf Published in Phys. Rev. B {\bf 66}, 235414 (2002)} 
\end{center}

\begin{abstract}
The vibrational properties (phonons) of a one-dimensional periodic phase
of atoms physisorbed in
the external groove of the carbon nanotube bundle are
studied. Analytical expressions for the phonon dispersion relations are
derived. The derived expressions are applied to Xe, Kr and Ar
adsorbates. The specific heat pertaining to dense phases of these
adsorbates is calculated.
\end{abstract}


\begin{center}
I. INTRODUCTION
\end{center}

Adsorption of gases in nanotube materials has been investigated
extensively in recent years, both from the experimental
\cite{Teizer,MigoneXe,MigoneCh4,MigonePRL,Hone,Las1} and theoretical
\cite{uptake,cylind,Siber1,Siber2,Gatica1,Colpriv,ColeCol}
standpoints. A very special geometry of the nanotube materials provides a
possibility for realization of unique adsorbate phases with reduced
effective dimensionality \cite{Siber1,Siber2,ColeCol,ColePRL}. The
behavior of adsorbates depends on whether they
are adsorbed within the tubes\cite{cylind}, in the interstitial
channels\cite{Siber1},
or in the
grooves.\cite{Siber2,Gatica1} For samples consisting of nanotubes with
closed ends, the
adsorption in the tube interior is not possible, and for sufficiently
large adsorbates, the adsorption potential in the interstitial channel is
purely repulsive.\cite{uptake} Thus, for the case of large atoms adsorbed 
in the material made of closed-end carbon nanotubes, the only positions
available for gas adsorption are the grooves of the bundles of carbon
nanotubes.

In this paper we consider a dense, periodic phase of
noble gas atoms physisorbed in the groove positions of a nanotube
bundle. This phase can
be visualized as a one-dimensional chain of atoms arranged in a periodic
array within a groove (see Fig. \ref{fig:fig1}). While the single
particle properties of quantum adsorbates
in grooves have been considered in Ref. \onlinecite{Siber2}, in
this article we concentrate on the many-particle excitations (phonons) of
the dense, periodic adsorbate
phase, that crucially depend on the interactions between
adsorbates. 

Recently, measurements of the specific heat of atoms adsorbed in nanotube
materials have been reported \cite{Hone,Las1}. In this respect, it is
important to theoretically consider the specific heat of a dense phase and
its signature in the overall specific heat of the sample (nanotube
material + adsorbates). 

The outline of the article is as follows. In Sec. II, we present a
simple lattice dynamics approach to the vibrations of adsorbates in
a groove. We show how the phonon frequencies are related to the
interadsorbate interaction potential and the
interaction of an adsorbate with the surrounding nanotube
medium. In Sec. III, we
apply the model from Sec. II to adsorption of Xe, Kr and Ar atoms and we
calculate the characteristic vibrational frequencies pertaining to these
adsorbates. In Sec. IV, we use this information to
calculate the specific heat of dense adsorbate phases. Sec. V summarizes
the main results.

\begin{center}
II. VIBRATIONS OF DENSE PHASES OF ADSORBATES IN THE GROOVES
\end{center}

To study the vibrations of the adsorbates in a groove we
introduce a simplifying assumption of rigid substrate i.e. we do not
consider phonons of the nanotube material. In this case, the total
potential energy $\Phi$ of the adsorbate system is given as
\begin{equation}
\Phi = \frac{1}{2} \sum _{l,l'} ^{l \ne l'} v({\bf r}_l - {\bf r}_{l'}) +
\sum_{l} V({\bf r}_l).
\label{eq:totpot}
\end{equation}
Here, $v({\bf r}_l - {\bf r}_{l'})$ represents the interaction between the
adsorbates at ${\bf r}_l$ and ${\bf r}_{l'}$ positions. The indices $l$
and $l'$ denote the adsorbed atoms, and $V({\bf r}_l)$ represents the
interaction of the $l$-th adsorbate with the surrounding nanotube medium
(substrate). We assume that in the absence of vibrations, the adsorbates
are positioned in an infinitely long one-dimensional lattice (with the
lattice parameter $a$, see Fig. \ref{fig:fig1}) within a groove. We also
neglect all the interactions between the adsorbates which are not in the
same groove. This is an excellent approximation, due to a large diameter
of carbon nanotubes.

If the interaction with the substrate were vanishingly small,
the problem would reduce to three dimensional vibrations of an
unsupported chain of atoms. This problem can be easily solved analytically
as shown e.g. in Ref. \onlinecite{Wollappl}. Even with the external field
$V$, the problem can be analytically solved, following the lattice
dynamics approach. \cite{Wollappl,DeWette1,Huang} The eigen-frequencies of
the adsorbate system can be found from the matrix equation
\begin{equation}
[{\cal D}(q) - M\omega(q)^2] {\bf e}(q) = 0,
\label{eq:secular}
\end{equation}
where $M$ is the adsorbate mass, $\omega(q)$ is the vibrational frequency,
$q$ is the phonon wave vector parallel to the groove and ${\bf
e}(q)$ is the 3D phonon polarization vector. Note that the phonon wave
vector $q$ is a one-dimensional quantity since the adsorbate lattice
exhibits a 1D
translational invariance. The polarization vector ${\bf e}(q)$ is a
three-dimensional quantity since the displacements of adsorbate atoms in
all three spatial directions are allowed. 
If we take into account only pair interactions $v$ between the nearest
neighboring adsorbates and neglect the discrete nature of the tubes
surrounding a groove by effectively "smearing" the carbon atoms
along the tube surface \cite{uptake,cylind},
the 3 $\times$ 3 dynamical matrix, ${\cal D}(q)$ is given as
\begin{equation}
{\cal D} = 
        \left[ \begin{array}{ccc}
       2 \beta [1-\cos (qa)] + f_{xx} & f_{xy} & f_{xz} \\
       f_{xy} & 2 \alpha[1-\cos (qa)] + f_{yy} & f_{yz} \\
       f_{xz} & f_{yz} & 2 \alpha[1-\cos (qa)] + f_{zz} \end{array}
\right].
\label{eq:dynmat}
\end{equation}
The $x$-axis is oriented along the chain of adsorbates, $y$-axis passes
through the centers of the two tubes surrounding a groove, and the
$z$-axis in perpendicular to both $x$ and $y$ axes and points outward
from the bundle, as denoted in Fig. \ref{fig:fig1}. The external force
constants $f_{\mu, \nu}$ are given by
\begin{equation}
f_{\mu, \nu} = \left( \frac{\partial ^2 V}{\partial \mu \partial \nu}
\right)_0, \hspace{0.8mm} \mu, \nu = x,y,z,
\label{eq:extfc}
\end{equation}
where the subscript zero signifies that the derivatives should be taken at
the equilibrium position of the adsorbate.
As the tubes surrounding a groove are assumed to be smooth, the external
potential, $V({\bf r}_l)$ does not depend on the adsorbate $x$-coordinate
and there is no external (substrate induced) restoring force associated
with the adsorbate
displacements in $x$-direction. That is why all $f_{\mu, \nu}$ constants
with $\mu=x$ or
$\nu=x$ vanish in Eq. (\ref{eq:dynmat}). Additionally, due to the symmetry
of the external potential with respect to the displacements parallel to
the line joining the centers of the two tubes surrounding a groove
($y$-direction, see Fig. \ref{fig:fig1}), the constant $f_{yz}$ also
vanishes.

The force constants associated
with the adsorbate-adsorbate interactions [$\alpha$ and $\beta$ in
Eq. (\ref{eq:dynmat})] are given as
\begin{eqnarray}
\alpha&=&\frac{1}{a} \left( \frac{dv}{dr} \right)_{r=a} \nonumber \\
\beta &=& \left( \frac{d^2 v}{d^2 r} \right)_{r=a},
\label{eq:alphbet}
\end{eqnarray}
where $r$ is the coordinate of relative distance between the
neighboring adsorbates. The phonon dispersion relations, $\omega(q)$ for
$-\pi/a \le
q \le \pi/a$, obtained as
the solutions to Eq. (\ref{eq:secular}) with
$f_{xx}=f_{xy}=f_{xz}=f_{yz}=0$ are given as
\begin{eqnarray}
\omega_{L}(q) &=& 2 \sqrt{\frac{\beta}{M}} \sin \left( \frac{qa}{2}
\right) \nonumber \\ 
\omega_{T1}(q) &=& \sqrt{\frac{f_{yy} + 2 \alpha [1-\cos(qa)]}{M}}
\nonumber
\\
\omega_{T2}(q) &=& \sqrt{\frac{f_{zz} + 2 \alpha [1-\cos(qa)]}{M}}.
\label{eq:disper}
\end{eqnarray}
where the subscripts $L,T1,T2$ denote the longitudinal ($L$) and the two
transverse phonon branches ($T1, T2$). The $L,T1$ and $T2$ modes are
polarized exclusively in $x,y$, and $z$ directions, respectively. Similar
equations have been derived in Ref. \onlinecite{Colpriv}. The authors of
this reference considered an isotropic oscillator model,
$V(x,y,z)=\overline{f} [(y-y_0)^2+(z-z_0)^2]/2$. The two transverse modes
are degenerate in this approximation, which however is not always the
case, because generally $f_{yy} \ne f_{zz}$, as will be shown in the next
section.

A brief
comment concerning the neglect of corrugation of the holding potential,
i.e. discreteness of the nanotube, is
in order here. If the adsorbate phase is commensurate with the
holding potential provided by the substrate, the longitudinal mode will
exhibit a zone center gap which can be related to the magnitude of
the corrugation. This effect has been experimentally confirmed for Xe
overlayers on Cu(111) surface \cite{SiberXe,SiberPRL} and should in
principle be observable for all commensurate systems.

\begin{center}
III. VIBRATIONS OF Xe, Kr AND Ar ADSORBATES
\end{center}

The ingredients necessary for the calculation of phonon modes are
obviously the force constants $\alpha, \beta$ and $f_{yy}, f_{zz}$.
The first two can be calculated
from the presumably known interadsorbate interaction potential, while
the $f$-force constants can be calculated from the asdsorbate-substrate
interaction potential. 

The adsorbate-substrate interaction can be calculated as a superposition
of the adsorbate-nanotube interactions. It is sufficient to consider
only the two nanotubes surrounding the groove, since the interaction of
adsorbate atom with other nanotubes is vanishingly small due to a large
diameter of the carbon nanotubes typically encountered in
experiments. \cite{Thess} Neglecting the discrete nature of the carbon
nanotube, the adsorbate-single wall nanotube interaction can be written
as \cite{uptake,cylind}
\begin{equation}
V(\rho) = 3 \pi \theta \epsilon \sigma^2 \left [ \frac{21}{32} \left(
\frac{\sigma}{R} \right )^{10} \eta ^{11} M_{11}(\eta) - \left(
\frac{\sigma}{R} \right )^{4} \eta ^{5} M_{5}(\eta) \right ],
\label{eq:tubepot}
\end{equation}
where $\epsilon$ and $\sigma$ are the energy and range parameter of the
effective adsorbate-carbon site interaction which is assumed to be of a
Lennard-Jones form. The variable $\rho$ denotes the distance of the
adsorbate atom
from the axis of the tube, $\theta$ is the effective coverage of C atoms
on the tube surface ($\theta$=0.38 1/\AA$^2$), $R$ is the tube radius and
the variable $\eta$ is defined as $\eta=R/\rho$ when
$\rho>R$. The function
$M_n(\eta)$ is
defined as
\begin{equation}
M_n(\eta) = \int_{0}^{\pi} \frac{d \phi}{ (1+\eta ^2-2
\eta \cos\phi) ^{n/2}}.
\end{equation}
The external potential can be constructed as a sum of the adsorbate
interactions with two tubes surrounding a groove. From the thus
constructed potential, the relevant external force constants can be
obtained using Eq. (\ref{eq:extfc}). 

To illustrate the
shape of the holding external potential, we plot in Fig. \ref{fig:fig2}
the potential for Xe adsorbate along the $y$ and
$z$ directions. In this figure and in all subsequent calculations, we
consider single wall carbon nanotubes with the diameter of 13.8 \AA
\hspace{0.7mm}, 
arranged within a bundle in a triangular lattice with the lattice constant
of 17 \AA.\cite{Thess}
The origin of the coordinate system is chosen in such a way that the
centers of the two tubes surrounding a groove are positioned at
$(y=0$ \AA, $z=0$ \AA) and ($y=17.0$ \AA, $z=0$ \AA).
The parameters of the Xe-C site potential are obtained
from the so-called combination rules, as explained in
Ref. \onlinecite{uptake}. We adopt all the relevant parameters of
adsorbate-adsorbate and adsorbate-substrate interactions from Table I of 
Ref. \onlinecite{uptake}. Since the two curves displayed in panels (a) and
(b) of Fig. \ref{fig:fig2} pass through the absolute minimum of the
external potential ($y=8.5$ \AA, $z=6.38$ \AA), the external force
constants can be obtained as second derivatives of the potential curves
presented in Fig. \ref{fig:fig2} at their minimum positions. Note that the
potential is significantly more anharmonic in the $z$ than in the $y$
direction which means that the harmonic approximation to the potential
is less satisfactory in the $z$-direction as can be seen from
inspection of thin dashed
lines in Fig. \ref{fig:fig2} which represent the harmonic
approximation. It should be mentioned that the minimum value of
the potential we
obtain is consistent with the one reported in
Ref. \onlinecite{uptake}. However, the binding energy of a single Xe
adsorbate
in this potential is not consistent with the experimental result
\cite{MigoneXe} (-282
$\pm$ 11 meV) which is lower than the absolute minimum of
the potential we obtain here (-225 meV). A possible small difference in
the tube separations at the surface of the bundle with respect to the
corresponding separation in the bundle interior may be
responsible for this discrepancy. In addition, as shown and discussed in
Ref. \onlinecite{Siber2}, small changes in the adsorbate-carbon site
effective potential may cause relatively large change in the adsorbate
binding energy.

The total adsorbate-adsorbate interaction encompasses direct
adsorbate-adsorbate interaction and the indirect interaction mediated by
the polarizable substrate material and other
adsorbates. \cite{Bruchbook,Klein1} This interaction is generally
different from the corresponding interaction between the atoms
in the gas phase. The most critical adsorbates to be considered in the
remainder of this paper are Xe adatoms which have the highest
polarizability of all the adsorbates we shall consider. While the
substrate mediated forces can be calculated for a
planar substrate material, it is more difficult to calculate them and
assess their importance for the
substrate of present interest, i.e. a bundle of carbon
nanotubes. This issue has been discussed in the case of interstitial
adsorption in Ref. \onlinecite{MKKost}.

Even though the many body forces should play a
role in dynamics of Xe monolayers on crystal surfaces\cite{Bruchbook}, it
is often found
that the gas phase Xe-Xe potentials provide very reliable phonon
dispersions \cite{BruchMolPhys,Bruchbook}. This is in part due to
the fact that phonon dispersions are not determined by the potential
itself, but by its derivatives which may be less influenced by many-body
effects. The only exception to this "rule"
seems to be the somewhat puzzling dynamics of Xe overlayers on Cu
surfaces. \cite{SiberXe,SiberPRL,APGrahCu,Bruchcomm,APreply,Wollnew} In
all the calculations to be presented, we shall assume that the
adsorbate-adsorbate interactions can be well represented by the
corresponding gas phase potentials.

We shall model the adsorbate-adsorbate interaction also by the
Lennard-Jones potential, $4 \epsilon_{gg} [ (\sigma_{gg}/r)^{12} -
(\sigma_{gg}/r)^{6}]$, where $r$ is the interadsorbate separation, and 
$\epsilon_{gg}$ and $\sigma_{gg}$ are the energy and
range parameters of the potential, respectively. The
equilibrium positions
of the adsorbates can be found from the static equilibrium condition,
i.e. by minimizing expression
(\ref{eq:totpot}) with
respect to the lattice parameter $a$. Following this procedure, we find
that the 1D lattice parameter is given by 
\begin{equation}
a= \sigma_{gg} \pi \left( \frac{1382}{675675} \right) ^{1/6} \approx 
1.1193 \sigma_{gg}.
\end{equation}
The force constants $\alpha$ and
$\beta$ [Eq. (\ref{eq:alphbet})] are then given as
\begin{eqnarray}
\alpha &=& -0.1665 \frac{\epsilon_{gg}}{\sigma_{gg}^2} \nonumber \\
\beta &=& 60.615 \frac{\epsilon_{gg}}{\sigma_{gg}^2}.
\end{eqnarray}
Force constant $\alpha$ determines the dispersion of the transverse
modes. It is negative due to the fact that the lattice parameter $a$ is
smaller than the equilibrium separation of the binary Lennard-Jones
potential (1.122 $\sigma_{gg}$). Note also that $| \alpha | \ll | \beta
|$, which means that the transverse modes have negligible dispersion in
comparison with the longitudinal mode. 

If the adsorbates form a chain that is commensurate with the substrate
corrugation, the periodicity of the one-dimensional chain of atoms may be
different from the one obtained here by neglecting the substrate
corrugation. In this case, all the force constants may change. On the
other hand, for the
incommensurate chain, our model should provide reliable results.

In Table I, we summarize the lattice parameters and the force
constants needed for calculation of the phonons in the specific cases of
Xe, Kr and Ar atoms adsorbed in a groove. We also summarize the mode
frequencies at the
Brillouin zone center ($q=0$) and the zone edge ($q=\pi/a$). As can be
seen from the table, the transverse modes for all adsorbates considered
have completely negligible dispersion throughout the Brillouin zone. We
also
find that $\omega_{T1}(q) < \omega_{T2}(q)$ which is a consequence of a
specific shape of the potential experienced by adsorbates in the
groove. This means that the potential is "softer" with respect to
adsorbate displacements in the $z$-direction (see Fig. \ref{fig:fig2}).

\begin{center}
IV. SPECIFIC HEAT OF DENSE ONE-DIMENSIONAL ADSORBATE PHASES
\end{center}

Knowing the dispersion relations for the adsorbate phonons, we can now
proceed to calculate the specific heat pertaining to a dense phase of
the adsorbates in a groove. The heat capacity at constant volume, $C_V$, 
for a collection of independent oscillators (phonons) is given by
\cite{Huang}
\begin{equation}
C_V = k_B \sum_{i} \frac{\left( \frac{\hbar \omega_i}{k_B T} \right) ^2
e^{\hbar \omega_i / k_B T}}{(e^{\hbar \omega_i / k_B T} - 1)^2},
\label{eq:cvgeneral}
\end{equation}
where $k_B$ is the Boltzmann constant, $T$ is the temperature, and index
$i$ counts the harmonic oscillators of frequencies $\omega _i$. In
our case,
the index $i$ is replaced by the phonon wave vector, $q$ and the phonon
branch index $s$ ($s=L,T1,T2$). Relation (\ref{eq:cvgeneral}) is thus
rewritten as
\begin{equation}
C_V=\frac{N k_B a}{\pi} \sum_s \int _{0} ^ {\pi / a}
\frac{\left[ \frac{\hbar \omega_s (q)}{k_B T} \right] ^2 e^{\hbar \omega_s
(q) / k_B T}}{[e^{\hbar \omega_s (q) / k_B T} - 1]^2} dq,
\label{eq:cvgroove}
\end{equation}
where we have transformed the sum over wave vectors into an integral as
$\sum_{q} \rightarrow (L_x / 2 \pi) \int_{-\pi / a} ^{\pi / a} dq$. Here,
$L_x$ is
the length of the adsorbate chain (or groove) and the number of adsorbates
within a particular groove is $N=L_x / a$. It is easy to check that in the
limit of high temperatures
($k_B T \gg \hbar \omega_s (q), \hspace{0.8mm} s=L,T1,T2$), the
specific heat in Eq. (\ref{eq:cvgroove}) reduces to
its classical value, $C_V = 3 N k_B $.

In Fig. \ref{fig:fig3} we plot the specific heat for Xe,Kr and Ar
adsorbates obtained from Eq. (\ref{eq:cvgroove}) by using information
on the adsorbate phonons summarized in Eq. (\ref{eq:disper}) and Table
I. As
can be seen, at high temperatures all the three curves tend to the
classical limit.

To separate the influence of transverse modes, we plot in
Fig. \ref{fig:fig4} the three different contributions to the specific heat
of a chain of Xe atoms,
i.e. in the sum in Eq. (\ref{eq:cvgroove}) we consider only one mode
(either $L, T1$, or $T2$). As can be inferred from this figure, the low
temperature behavior
of the specific heat ($T < 4$ K) is completely determined by the
longitudinal
mode. At temperatures higher than about 4 K (for Xe), all the modes in 
the sum in Eq. (\ref{eq:cvgroove}) must be considered.
Due to the fact that the
transverse modes have higher frequencies for Kr and Ar adsorbates, their
contribution to the specific heat becomes significant at somewhat higher
temperatures. The specific heat at low temperatures, where
only the L mode contributes, is proportional to the temperature and the
following relation holds:
\begin{equation}
C_V \approx 2.095 N k_{B} \frac{k_B T}{\hbar \omega_L (q=\pi/a)}.
\label{eq:linearcv}
\end{equation}
The linear dependence of specific heat on temperature is a consequence
of effectively 1D behavior of the chain at low temperatures and can be
easily seen from Fig. \ref{fig:fig3} at temperatures lower than $\approx$
4K. Equation (\ref{eq:linearcv}) also implies that the measurement of
specific heat at low temperatures would yield
an information on the maximum frequency of the longitudinal mode
(Brillouin zone edge frequency). As this frequency crucially depends on
the interadsorbate potential [Eq. (\ref{eq:alphbet}) and
Eq. (\ref{eq:disper})], these measurements could yield an information on
the effective interaction between the adsorbate atoms which may be
modified by the presence of the substrate material, as discussed in
Sec. III.

\begin{center}
V. SUMMARY
\end{center}

We have studied the vibrational properties of linear dense phases of
adsorbates in the grooves of the carbon nanotube bundles. The three
modes characterizing the adsorbate vibrations have been identified and
their frequencies have been calculated for Xe, Kr and Ar adsorbates. We
have also calculated the specific heat corresponding to the linear phase
and identified the temperature regimes in which the transverse modes begin
to significantly contribute to the specific heat. The presented results
should serve as a useful guidance in the measurements of specific heat of
gases adsorbed in the nanotube
materials. \cite{Hone,Las1}

\begin{center}
ACKNOWLEDGMENTS
\end{center}

I thank dr. Branko Gumhalter and Marko T. Cvita\v{s} for a careful reading
of the manuscript. I thank prof. Milton W. Cole for bringing
Ref. \onlinecite{Colpriv} to my attention.

\begin{figure}
\caption{
A sketch of a small bundle of carbon nanotubes with the one-dimensional
adsorbate phase in one of its grooves. The choice of the coordinate
system is denoted. The adsorbate atoms are denoted by small dark circles.}
\label{fig:fig1}
\end{figure}

\begin{figure}
\caption{
Full lines: External potential for Xe atom in the groove. (a) Potential
along the $z$-direction. (b) Potential along the $y$-direction. Both
curves pass through the absolute minimum of the potential located at
($y=8.5$ \AA, $z=6.38$ \AA). Dashed lines: Harmonic
approximation to the potential.}
\label{fig:fig2}
\end{figure}

\begin{figure}
\caption{
Specific heat of Xe (thick full line), Kr (thick dashed line) and Ar
(thick dotted line) adsorbate chains in a groove as a function of the
temperature. The thin full line represents the linear low-temperature
behavior of specific heat for Ar as predicted by Eq. (\ref{eq:linearcv}), 
with $\hbar \omega_{L}(q= \pi / a) = 4.93$ meV (see Table I.).
}
\label{fig:fig3}
\end{figure}

\begin{figure}
\caption{
Separate contributions of three modes of Xe chain to the specific heat.
Dashed line: $L$ mode. Dash-dotted line: $T2$ mode. Dash-dash-dotted
line: $T1$ mode. Full line: total specific heat of the Xe chain.}
\label{fig:fig4}
\end{figure}

\begin{table}
\begin{tabular}{|c|c|c|c|}
       & Xe & Kr & Ar \\
\hline
$a$ [\AA]     & 4.59  & 4.03  & 3.81  \\ 
\hline
$f_{yy}$ [N/m] & 7.13  & 6.50  & 5.58 \\
\hline
$f_{zz}$ [N/m] & 4.01  & 3.19  & 2.58 \\
\hline
$\alpha$ [N/m] & -0.003  & -0.003  & -0.0024  \\
\hline
$\beta$ [N/m] & 1.10  & 1.10  & 0.87  \\
\hline
$\hbar \omega_{L} (q=0)$ [meV] & 0 & 0 & 0 \\
\hline
$\hbar \omega_{L} (q=\pi/a)$ [meV] & 3.06 & 3.83 & 4.93 \\
\hline
$\hbar \omega_{T1} (q=0)$ [meV] & 3.899 & 4.660 & 6.253 \\
\hline
$\hbar \omega_{T1} (q=\pi/a)$ [meV] & 3.895 & 4.655 & 6.247 \\
\hline
$\hbar \omega_{T2} (q=0)$ [meV] & 2.924 & 3.264 & 4.251 \\
\hline
$\hbar \omega_{T2} (q=\pi/a)$ [meV] & 2.919 & 3.258 & 4.244 \\
\end{tabular}
\caption{Force constants and phonon frequencies of $L, T1$ and $T2$
modes (Brillouin zone center and Brillouin
zone edge) for Xe, Kr and Ar adsorbates in groove positions of the
carbon nanotube bundle.}
\end{table}

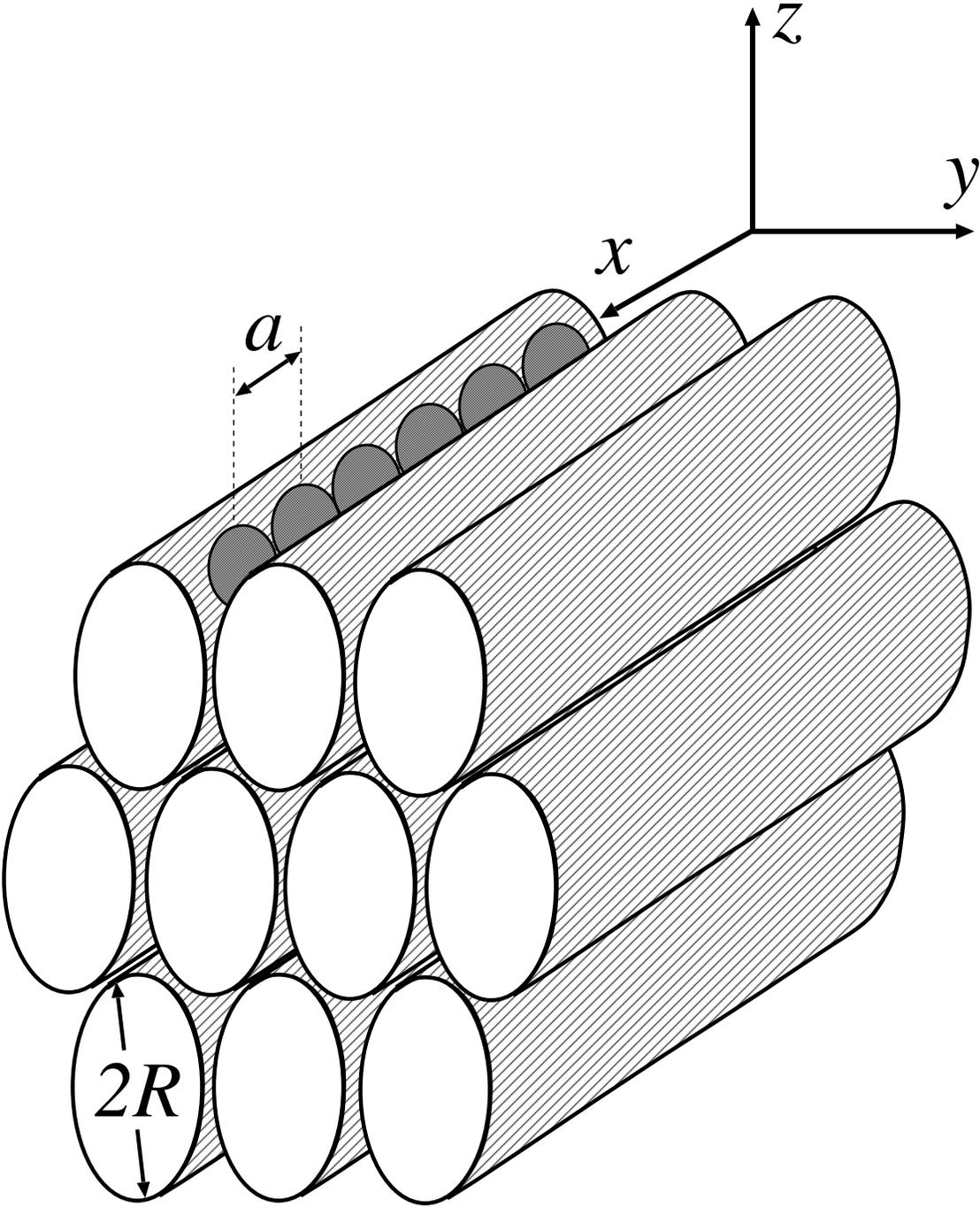
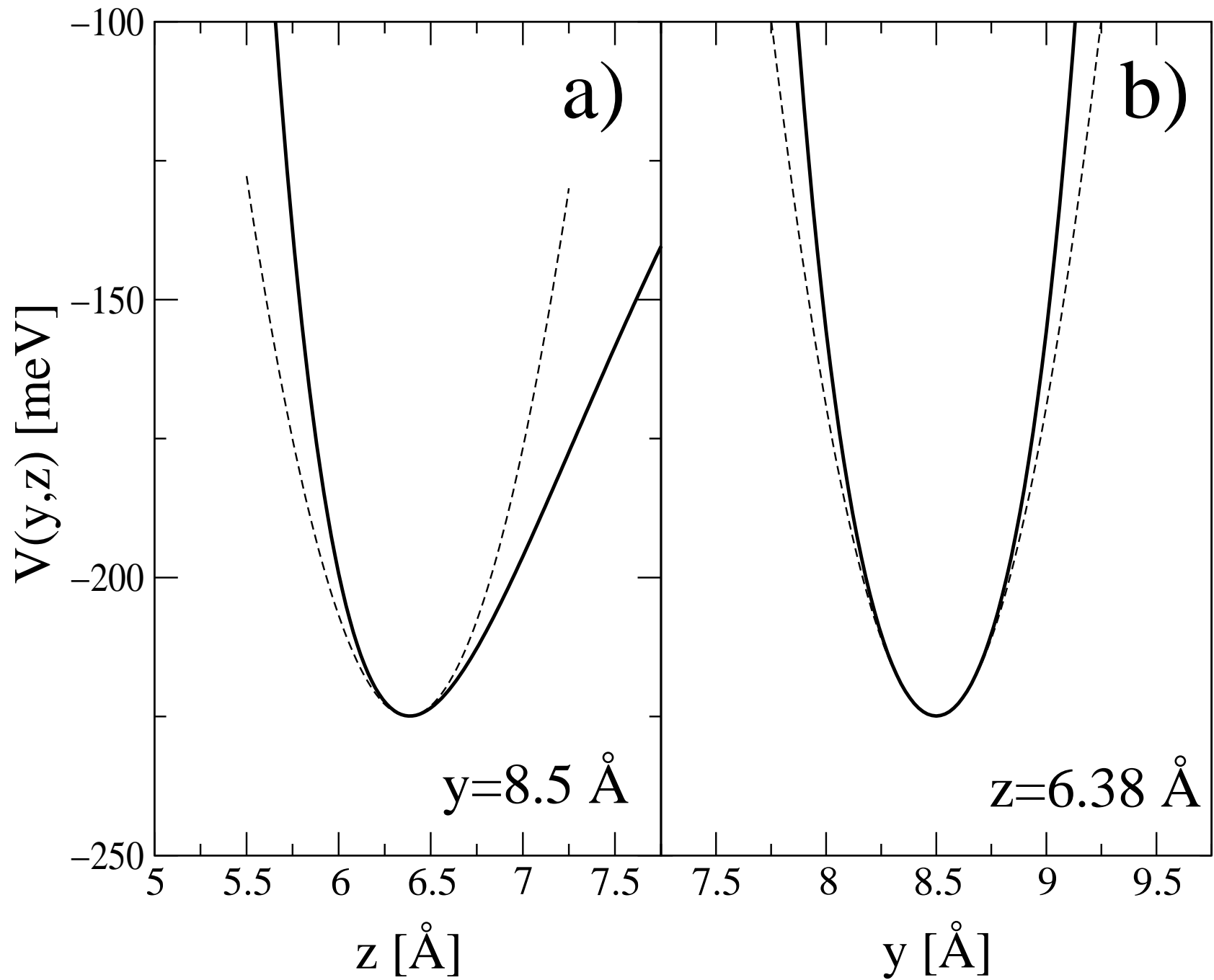
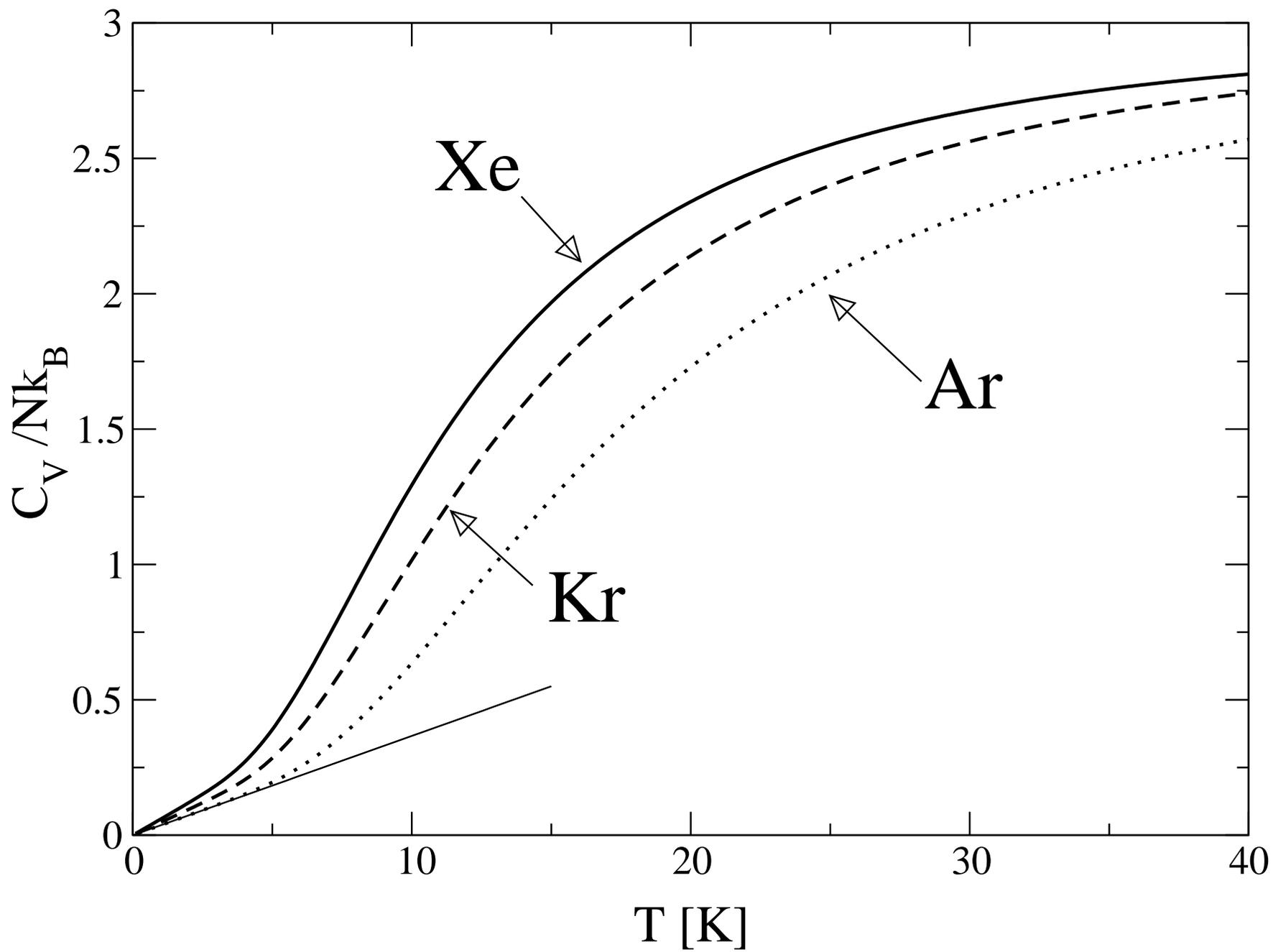
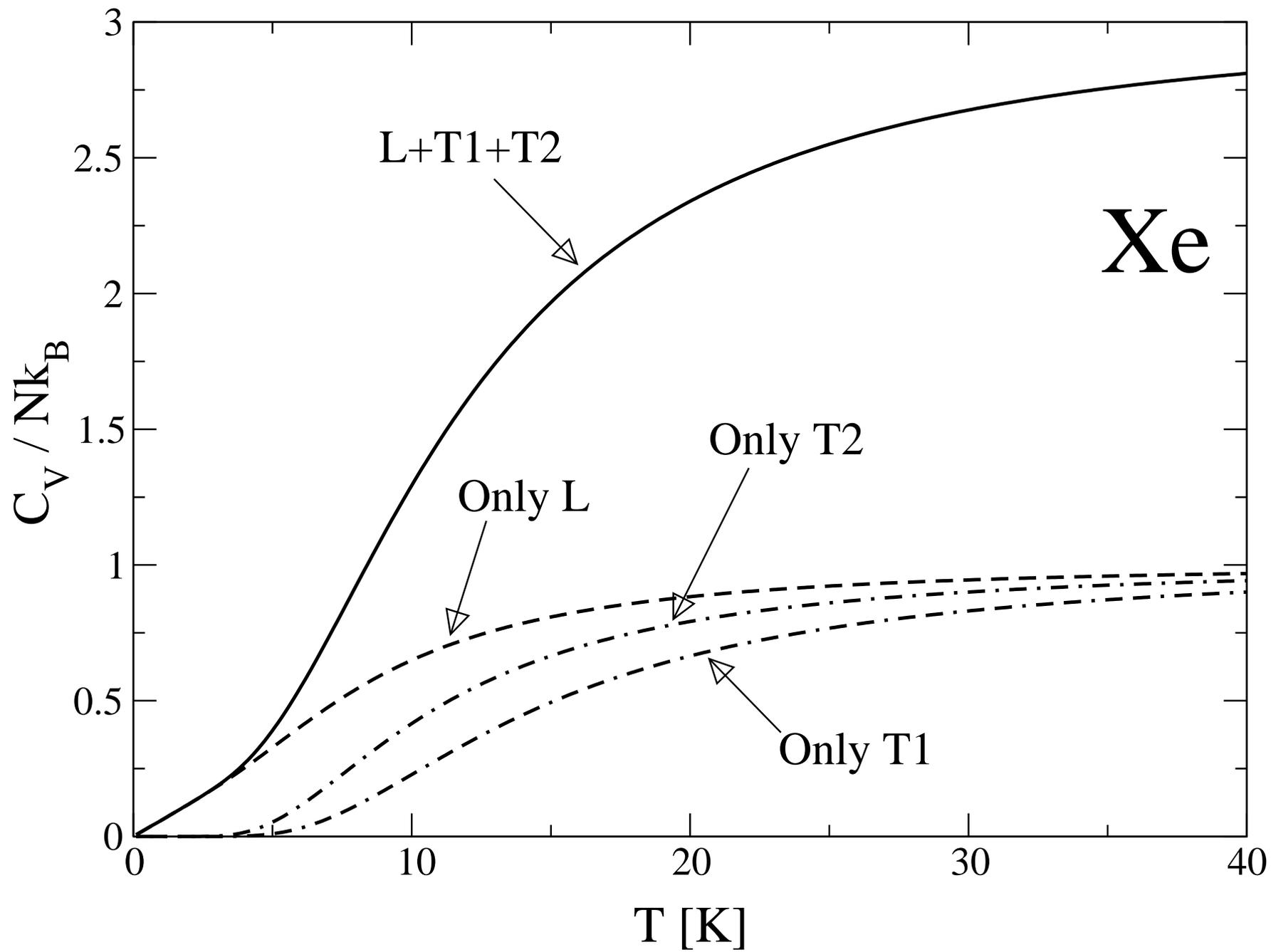
\end{document}